\documentstyle[12pt]{article}

\def\hybrid{\topmargin -20pt	\oddsidemargin 0pt
	\headheight 0pt	\headsep 0pt
        \textwidth 6.35in
        \textheight 9.65in
	\marginparwidth .875in
	\parskip 5pt plus 1pt	\jot = 1.5ex}

 \catcode`\@=11

\@addtoreset{equation}{subsection}
\@addtoreset{footnote}{section}
\def\theequation{\thesubsection.\arabic{equation}}

\newtoks\@stequation

\def\subequations{\refstepcounter{equation}%
  \edef\@savedequation{\the\c@equation}%
  \@stequation=\expandafter{\theequation}
  \edef\@savedtheequation{\the\@stequation}
  \edef\oldtheequation{\theequation}%
  \setcounter{equation}{0}%
  \def\theequation{\oldtheequation\alph{equation}}}

\def\endsubequations{\setcounter{equation}{\@savedequation}%
  \@stequation=\expandafter{\@savedtheequation}%
  \edef\theequation{\the\@stequation}\global\@ignoretrue
  \vspace*{-12pt} \\}

\def\np{Nucl. Phys. \/}
\def\pl{Phys. Lett. \/}

\hybrid
\def\baselinestretch{1.2}
\catcode`\@=11
\newcount\hour
\newcount\minute
\newtoks\amorpm
\hour=\time\divide\hour by60
\minute=\time{\multiply\hour by60 \global\advance\minute by-\hour}
\edef\standardtime{{\ifnum\hour<12 \global\amorpm={am}%
	\else\global\amorpm={pm}\advance\hour by-12 \fi
	\ifnum\hour=0 \hour=12 \fi
	\number\hour:\ifnum\minute<10 0\fi\number\minute\the\amorpm}}
\edef\militarytime{\number\hour:\ifnum\minute<10 0\fi\number\minute}

\def\draftlabel#1{{\@bsphack\if@filesw {\let\thepage\relax
   \xdef\@gtempa{\write\@auxout{\string
      \newlabel{#1}{{\@currentlabel}{\thepage}}}}}\@gtempa
   \if@nobreak \ifvmode\nobreak\fi\fi\fi\@esphack}
	\gdef\@eqnlabel{#1}}
\def\@eqnlabel{}
\def\@vacuum{}
\def\draftmarginnote#1{\marginpar{\raggedright\scriptsize\tt#1}}

\def\draft{\oddsidemargin -.5truein
	\def\@oddfoot{\sl preliminary draft \hfil
	\rm\thepage\hfil\sl\today\quad\militarytime}
	\let\@evenfoot\@oddfoot	\overfullrule 3pt
	\let\label=\draftlabel
	\let\marginnote=\draftmarginnote
   \def\@eqnnum{(\theequation)\rlap{\kern\marginparsep\tt\@eqnlabel}%
\global\let\@eqnlabel\@vacuum}  }

\catcode`\@=11
\def\section{\@startsection {section}{1}{0pt}{-3.5ex plus -1ex minus
 -.2ex}{2.3ex plus .2ex}{\raggedright\large\bf}}
\catcode`\@=12
\newskip\humongous \humongous=0pt plus 1000pt minus 1000pt

\newif\ifdtup

\def\be{\begin{equation}}
\def\ee{\end{equation}}
\def\ba{\begin{eqnarray}}
\def\ea{\end{eqnarray}}
\def\bs{\begin{subequations}}
\def\es{\end{subequations}}

\def\RR{{\rm I\!R}}
\def\R{{\cal{R}}}
\def\p{\partial}
\def\pq{\bar {\cal P}}

\def\pb{\bar \partial}
\def\a{\alpha}
\def\b{\beta}
\def\g{\gamma}
\def\n{\nabla}

\def\np{Nucl. Phys. \/}
\def\pl{Phys. Lett. \/}

\def\Q{{\cal{Q}}}
\begin{document}
\renewcommand{\theequation}{\thesection.\arabic{equation}}
\newcommand{\beq}{\begin{equation}}
\newcommand{\eeq}[1]{\label{#1}\end{equation}}
\newcommand{\ber}{\begin{eqnarray}}
\newcommand{\eer}[1]{\label{#1}\end{eqnarray}}
\begin{titlepage}
\begin{center}

\hfill CERN-TH/95-223\\
\hfill LPTENS-95/41\\
\hfill hep-th/9509043\\

\vskip .2in

{\large \bf Instabilities in Strong Magnetic Fields in String
Theory\footnote{To appear in the proceedings of the Conference on
Gauge Theories, Applied Supersymmetry and Quantum Gravity, Leuven,
July 1995.}}
\vskip .4in

{\bf Elias Kiritsis and Costas Kounnas\footnote{On leave from Ecole
Normale Sup\'erieure, 24 rue Lhomond, F-75231, Paris, Cedex 05,
FRANCE.}}\\
\vskip
 .3in

{\em Theory Division, CERN,\\ CH-1211,
Geneva 23, SWITZERLAND} \\

\vskip .3in

\end{center}

\vskip .2in

\begin{center} {\bf ABSTRACT } \end{center}
\begin{quotation}\noindent
We construct groundstates of the string with non-zero mass gap and
non-trivial chromo-magnetic fields as well as
curvature.
The exact spectrum as function of the chromo-magnetic fields and
curvature is derived.
We examine the behavior of the spectrum, and find that there is a
maximal value for the magnetic field $H_{\rm max}\sim M_{\rm
Plank}^2$. At
this value all states that couple to the magnetic field become
infinitely massive and decouple.
We also find tachyonic
instabilities for
strong background fields of the order ${\cal O}(\mu M_{\rm Plank})$
where $\mu$ is the mass gap of the theory. Unlike the field theory
case, we find that such ground states become stable again for
magnetic fields
of the order ${\cal O}(M^2_{\rm Plank})$.
The implications of these results are discussed.

\end{quotation}
\vskip 3.0cm
CERN-TH/95-223\\
September 1995\\
\end{titlepage}
\vfill
\eject
\def\baselinestretch{1.2}
\baselineskip 16 pt
\noindent
\section{Introduction}
\setcounter{equation}{0}
In  four-dimensional Heterotic or type II  Superstrings it is
possible, in principle, to understand the response of the theory  to
non-zero  gauge or  gravitational field  backgrounds including
quantum corrections.
This problem is difficult in its full generality since
we are working in a first quantized framework. In certain special
cases, however, there is an underlying 2-d superconformal theory
which
is well understood and which describes exactly (via marginal
deformations) the turning-on of non-trivial
gauge and gravitational backgrounds.
This exact description goes beyond the linearized approximation.
In such cases, the spectrum can be calculated exactly, and it can
provide
interesting clues about the physics of the theory.

In field theory (excluding gravity) the
energy shifts of a state due to the magnetic field have been
investigated long ago \cite{field,qcd,stand}. The classical field
theory
formula for the energy of a state with spin $S$, mass $M$ and charge
$e$ in a magnetic field
$H$ pointing in the third direction is:
\be
E^2=p_3^2+M^2+|eH|(2n+1-gS)\label{class}
\ee
were $g=1/S$ for minimally coupled states and $n=1,2,...$ labels the
Landau levels.
It is obvious from (\ref{class}) that minimally coupled particles
cannot become tachyonic, so the theory is stable.
For non-minimally coupled particles, however, the factor $2n+1-gS$
can become negative and instabilities thus appear.
For example, in non-abelian gauge theories, there are particles which
are not minimally coupled. In the standard model, the $W^{\pm}$
bosons
have $g=2$ and $S=\pm 1$. From (\ref{class}) we obtain that the
spontaneously broken phase in the standard model is thus unstable for
 magnetic fields that satisfy \cite{qcd,stand}
\be
|H|\geq {M_{W}^2\over |e|}  \label{sta}
\ee
A phase transition has to occur by a condensation of $W$ bosons, most
probably to a phase where the magnetic field is confined (localized)
in a tube, \cite{stand}.
This behavior should be contrasted to the constant electric field
case where
there is particle production \cite{sch} for any non-zero value of the
electric field, but the vacuum is stable (although the particle
emission tends to decrease the electric field).

The instabilities present for constant magnetic fields are still
present in general for slowly varying (long range) magnetic fields.
For a non-abelian gauge theory in the unbroken phase, since the mass
gap
is classically zero, we deduce from (\ref{class}) that the trivial
vacuum
($A_{\mu}^{a}=0$) is unstable even for infinitesimally small
chromo-magnetic
fields. This provides already an indication at the classical level
that the trivial vacuum is not the correct vacuum in an unbroken
non-abelian gauge theory. We know however, that such a theory
acquires a non-perturbative mass gap, $\Lambda^2\sim \mu^2
\exp[-16\pi^2 b_{0}/g^2]$ where $g$ is the non-abelian gauge
coupling.
If in such a
theory one managed to create a chromo-magnetic field then there would
again appear an instability and the theory would again confine the
field in a flux tube.

In string theory, non-minimal gauge couplings are present not only in
the massless sector but also in the massive (stringy) sectors
\cite{fpt}.
Thus one would expect similar instabilities.
Since in  string theory there are states with arbitrary large values
of spin
and one can naively expect that if  $g$ does not decrease fast enough
with the spin
(as is the case in open strings where $g=2$ \cite{fpt}) then for
states with
large spin an arbitrarily small magnetic field would destabilize the
theory. This behavior would imply that the trivial vacuum is
unstable.
This does not happen however since the masses of particles with spin
also become large when the spin gets large.

The spectrum of open bosonic strings in constant magnetic fields was
derived in \cite{acny}. Open bosonic strings however, contain
tachyons even in the absence
of background fields. It is thus more interesting to investigate
open superstrings which are tachyon-free.
This was done in  \cite{fp}.
It was found that for weak magnetic fields the field theory formula
(\ref{class}) is obtained, and there are similar instabilities.

In closed superstring theory however, one is forced to include the
effects of gravity.
A constant magnetic field for example carries energy, thus, the space
cannot remain flat anymore.
The interesting question in this context is, to what extend, the
gravitational backreaction changes the behavior seen in field theory
and open string theory.
As we will see the gravitational backreaction is important and gives
rise to interesting new phenomena in strong magnetic fields.

Such questions can have potential interesting applications in string
cosmology
since long range magnetic fields can be produced at early stages in
the history of the universe where field theoretic behavior can be
quite different from the stringy one.

The first example of an exact electromagnetic solution to closed
string theory
was described in \cite{bas}.
The solution included both an electric and magnetic field
(corresponding to the electrovac solution of supergravity).
In \cite{bk} another exact closed string solution was presented
(among others)
which corresponded to a Dirac monopole over $S^2$.
More recently, several other magnetic solutions were presented
corresponding to
localized \cite{melvin} or covariantly constant magnetic fields
\cite{rt}.
The spectrum of these magnetic solutions seems to have a different
behavior
as a function of the magnetic field, compared to the situation
treated in this paper. The reason for this is that \cite{rt}
considered magnetic solutions where the gravitational backreaction
produces a non-static metric.
``Internal" magnetic fields of the type described in \cite{bk} were
also considered recently \cite{b} in order to break spacetime
supersymmetry.

Here we will study  the effects of  covariantly  constant
(chromo)magnetic fields, $H^a_i=\epsilon^{ijk}F_{jk}^a$ and constant
curvature
${\cal R}^{il}=\epsilon^{ijk}\epsilon^{lmn}{\cal R}_{jm,kn}$, in
four-dimensional closed superstrings.
The relevant framework was developed in \cite{kk} where ground
states were
found, with a continuous (almost constant) magnetic field in a weakly
curved space.
We will describe here the relevant framework and physics of such
backgrounds.
More details and conventions can be found in \cite{magnetic}.

In the heterotic string (where the left moving
sector is N=1 supersymmetric) the part of the $\sigma$-model action
which corresponds to a gauge field background $A^{a}_{\mu}(x)$
is
\be
V=(A^{a}_{\mu}(x)\partial x^{\mu}+F^{a}_{ij}(x)\psi^{i}\psi^{j})\bar
J^{a}
\label{vertex}
\ee
where $F^{a}_{ij}$ is the field strength of $A^{a}_{\mu}$ with
tangent
space indices, eg. $F^{a}_{ij}=e^{\mu}_{i}e^{\nu}_{j}F^{a}_{\mu\nu}$
with $e^{\mu}_{i}$ being the inverse vielbein, and $\psi^{i}$ are
left-moving world-sheet fermions with a normalized kinetic term.
$\bar J^{a}$ is a right moving affine current.

Consider a string ground state with a flat non-compact (euclidean)
spacetime ($\RR^{4}$).
The simplest case to consider is that of a constant magnetic field,
$H^a_i=\epsilon^{ijk}F_{jk}^a$.
Then the relevant vertex operator (\ref{vertex}) becomes
\be
V_{flat}=F_{ij}^a({1\over 2}x^{i} \partial
x^{j}+\psi^{i}\psi^{j})\bar
J^{a}
\label{vertexflat}
\ee
This vertex operator however, cannot be used to turn on the magnetic
field
since it is not marginal (when $F_{ij}^{a}$ is constant). In other
words, a constant magnetic field in flat
space
does not satisfy the string equations of motion, in particular the
ones
associated with the gravitational sector.

A way to bypass this problem we need to switch on more background
fields.
In \cite{kk} we achieved this in two steps.
First, we found an exact string ground state in which $\RR^{4}$ is
replaced by
$\RR\times S^{3}$. The $\RR$ part corresponds to free boson with
background charge
$Q=1/\sqrt{k+2}$ while the $S^{3}$ part corresponds to an $SU(2)_{k}$
WZW model. For any (positive integer) k, the combined central charge
is equal to that of $\RR^{4}$. For large $k$, this background has a
linear dilaton in the $x^{0}$ direction as well as an
$SO(3)$-symmetric antisymmetric tensor on $S^{3}$, while the metric
is
the standard round metric on $S^{3}$ with constant
curvature.
On this space, there is an exactly marginal vertex operator for a
magnetic field which is
\be
V_{m}=H(J^{3}+\psi^{1}\psi^{2})\bar J^{a}\label{magnet}
\ee
Here, $J^{3}$ is the left-moving current of the $SU(2)_{k}$ WZW
model.
$V_{m}$ contains the only linear combination of $J^{3}$ and
$\psi^{1}\psi^{2}$
that does not break the N=1 local supersymmetry.
The exact marginality of this vertex operator is obvious since it is
a product of a left times a right abelian current.
This operator is unique up to an $SU(2)_{L}$ rotation.

We can observe  that this vertex operator provides a well defined
analog
of $V_{flat}$ in eq. (\ref{vertex}) by looking at the large $k$
limit.
We will write the SU(2) group element as
$g=\exp[i{\vec\sigma}\cdot{\vec x}/2]$ in which case
$J^{i}=kTr[\sigma^{i}
g^{-1}\p g]=ik(\p x^{i}+\epsilon^{ijk}x_j\p x_k+{\cal{O}}(|x|^3))$.
In the flat limit the first term corresponds to a constant gauge
field
and thus pure gauge so the only relevant term is the second one which
corresponds to constant magnetic field in flat space.
The fact  $\pi_{2}(S^{3})=0$ explains in a different way why there is
no quantization condition on $H$.

There is another exactly marginal perturbation in the background
above
that turns on fields in the gravitational sector.
The relevant perturbation is
\be
V_{grav}=\R (J^{3}+\psi^{1}\psi^{2})\bar J^{3}
\label{gravi}\ee
This perturbation modifies the metric, antisymmetric tensor and
dilaton \cite{kk}.
For type II strings the relevant perturbation is
\be
V_{grav}^{II}=\R(J^3+\psi^{1}\psi^{2})(\bar J^{3}+\bar\psi^{1}\bar
\psi^{2})
\ee
We will not describe this perturbation further. They have been
studied using the results of \cite{gk} and  we refer the interested
reader to \cite{magnetic} for more.

The space we are using, $\RR\times S^3$ is such that the spectrum has
a mass gap
$\mu^2$.
In particular all gauge symmetries are broken spontaneously. This
breaking
however is not the standard breaking due to a constant expectation
value of a scalar but due to non-trivial expectation values of the
fields in the universal sector (graviton, antisymmetric tensor and
dilaton).

\section{Effective Field Theory Analysis}
\setcounter{equation}{0}
\noindent
The starting 4-d spacetime (we will use Euclidean signature here) is
described
by the $SO(3)_{k/2}\times \RR_{Q}$ CFT.
The heterotic $\sigma$-model that describes this space is\footnote{
In most formulae we set $\alpha'=1$ unless stated otherwise.}

\be
S_{4d}={k\over 4}{\bf I}_{SO(3)}(\a,\b,\g)
+{1\over 2\pi}\int d^2z \left[\p x^{0}\pb
x^{0}+\psi^{0}\pb\psi^{0}+\sum_{a=1}^{3}\psi^{a}\pb\psi^{a}\right]+
{Q\over 4\pi}\int \sqrt{g}R^{(2)}x^{0}\label{initi}
\ee
while the SU(2) action can be written in Euler angles as
\be
{\bf I}_{SO(3)}(\a,\b,\g)={1\over 2\pi}\int d^2
z\left[\p\a\pb\a+\p\b\pb\b
+\p\g\pb\g+2\cos\b\p\a\pb\g\right]
\ee
with $\b\in[0,\pi]$, $\a,\g\in[0,2\pi]$ and $k$ is a positive even
integer.
In the type II case we have to add also the right moving fermions
$\bar \psi^{i}$, $1\leq i\leq 4$. The fermions are free (this is a
property
valid for all supersymmetric WZW models).

Comparing with the general (bosonic) $\sigma$-model
\be
S={1\over 2\pi}\int d^2 z (G_{\mu\nu}+B_{\mu\nu})\p x^{\mu}\pb
x^{\nu}+
{1\over 4\pi}\int \sqrt{g}R^{(2)}\Phi(x)
\ee
we can identify the non-zero background fields as
\be
G_{00}=1\;\;,\;\; G_{\a\a}=G_{\b\b}=G_{\g\g}={k\over 4}\;\;,\;\;
G_{\a\g}={k\over 4}\cos\b\;\;\;,\;\;\;B_{\a\g}={k\over 4}\cos\b
\label{3s2}\ee
\be
\Phi=Qx^{0}={x^{0}\over \sqrt{k+2}}\label{dil}
\ee
where the relation between $Q$ and $k$ comes from the
requirement that
the (heterotic) central charge should be $(6,4)$, in which case we
have (4,0)
superconformal invariance, \cite{k}.

The perturbation that turns on a chromo-magnetic field in the $\mu=3$
direction
is proportional to $(J^{3}+\psi^1\psi^2)\bar J$ where $\bar J$ is a
right moving current belonging to the Cartan subalgebra of the
heterotic gauge group.
It is normalized so that $\langle\bar J(1)\bar J(0)\rangle=k_{g}/2$.
Since
\be
J^{3}=k(\p\g+\cos\b\p\a)\;\;\;,\;\;\;\bar J^{3}=k(\pb\a+\cos\b\pb\g)
\label{norm}
\ee
this perturbation changes the $\sigma$-model action in the following
way:
\be
\delta S_{4d}={\sqrt{kk_{g}}H\over 2\pi}\int d^2
z(\p\g+\cos\b\p\a)\bar J
\label{gauge}
\ee

In the type II case $\bar J$ is a bosonic current (it has a left
moving partner) and we can easily show that the $\sigma$-model with
action
$S_{4d}+\delta S_{4d}$ is conformally invariant to all orders in
$\alpha'$.

Reading the spacetime backgrounds from (\ref{initi}), (\ref{gauge})
is not entirely
trivial but straightforward.
In type II case (which corresponds to standard Kalutza-Klein
reduction) the correct metric  has an $A_{\mu}A_{\nu}$ term
subtracted \cite{dpn}.
In the heterotic case there is a similar subtraction but the reason
is different. It has to do with the anomaly in the holomorphic
factorization
of a boson.

The background fields have to be solutions (in leading order in $\a
'$) to equations of motion stemming from the following spacetime
action \cite{eff}:

\be
S=\int d^{4}x\sqrt{G}e^{-2\Phi}\left[R+4(\n\Phi)^2-{1\over
12}H^2-{1\over 4g^2}F^a_{\mu\nu}F^{a,\mu\nu}+{\delta c\over
3}\right]\label{5}
\ee
where we have displayed a  gauge field $A^{a}_{\mu}$, (abelian or
non-abelian) and set $g_{\rm string}=1$. The gauge coupling
is $g^2=2/k_{g}$ due to the normalization of the affine currents,
\be
H_{\mu\nu\rho}=\p_{\mu}B_{\nu\rho}-{1\over
2g^2}\left[A^a_{\mu}F^a_{\nu\rho}-{1\over
3}f^{abc}A^{a}_{\mu}A^{b}_{\nu}A^{c}_{\rho}\right]+{\rm
cyclic}\;\;{\rm permutations}\label{7}
\ee
and $f^{abc}$ are the structure constants of the gauge group.
In this paper we will restrict ourselves to abelian gauge fields (in
the cartan
of a non-abelian gauge group).

It is not difficult now to read from (\ref{initi}), (\ref{gauge}) the
background
fields
that satisfy the effective action equations.
The non-zero components are:
\be
G_{00}=1\;\;,\;\;G_{\b\b}={k\over 4}\;\;,\;\;G_{\a\g}={k\over
4}(1-2H^2)\cos\b\;\;,\;\;G_{\a\a}={k\over 4}(1-2H^2\cos^2\b)
\label{8}\ee
\be
G_{\g\g}={k\over
4}(1-2H^2)\;\;,\;\;B_{\a\g}={k\over
4}\cos\b\;\;,\;\;A_{a}=g\sqrt{k}H\cos\b\;\;\;,\;\;\;
A_{\g}=g\sqrt{k}H
\label{9}\ee
and the same dilaton as in (\ref{dil}).
This background is exact to all orders in the $\a '$
expansion
with the simple modification $k\to k+2$.

It is interesting to note that
\be
\sqrt{{\rm det}G}=\sqrt{1-2H^2}\left({k\over 4}\right)^{3/2}\sin\b
\ee
which indicates, as advertised earlier, that something special
happens
at $H_{max}=1/\sqrt{2}$.
At this point the curvature is regular.
In fact, this is a boundary point where the states that couple to the
magnetic field (i.e. states with non-zero left helicity and angular
momentum  and/or $e$) become
infinitely massive and decouple.
This is the same phenomenon as the degeneration of the Kh\"aler
structure on a two-torus (${\rm Im}U\to\infty$).
Thus, this point is at the boundary of the magnetic field moduli
space.
This is very interesting since it implies the existence of a maximal
magnetic
field
\be
|H|\leq H_{max}={1\over \sqrt{2}}\;\;\;\;{\rm or}\;\;\;\;H_{\rm
max}={M_{\rm Plank}^2\over \sqrt{2}}\label{max}
\ee
in physical units with $M^2_{\rm Plank}=1/\alpha'g^{2}_{\rm string}$
and where $g_{\rm
string}$ is the string coupling constant.

We should note here that the deformation of the spherical geometry by
the magnetic field is smooth for all ranges of parameters, even at
the boundary point $H=1/\sqrt{2}$.
To monitor better the back-reaction of the effective field theory
geometry
we should first write the three-sphere with the round metric
(\ref{3s2}), as the (Hopf)
fibration with $S^1$ as fiber and a two-sphere as base space:
\be
ds^2_{\rm 3-sphere}={k\over 4}\left[ds^2_{\rm 2-sphere}+(d\g+\cos\b
d\a)^2\right]\;\;,\;\;ds^2_{\rm 2-sphere}=d\b^2+\sin^2\b d\a^2
\label{hopf}
\ee
The second term in (\ref{hopf}) is the metric of the $S^1$ fiber, and
its non-trivial dependence on $\a,\b$ signals the non-triviality of
the Hopf fibration.
This metric has $SO(3)\times SO(3)$ symmetry.

The metric (\ref{8}), (\ref{9}) containing the backreaction to the
non-zero magnetic field can be written as
\be
ds^2={k\over 4}\left[ds^2_{\rm 2-sphere}+(1-2H^2)(d\g+\cos\b
d\a)^2\right]
\label{back}\ee
It is obvious from (\ref{back}) that the magnetic field changes the
radius
of the fiber and breaks the $SO(3)\times SO(3)$ symmetry to the
diagonal $SO(3)$.
It is also obvious that at $H=1/\sqrt{2}$, the radius of the fiber
becomes
zero.
All the curvature invariants are smooth (and constant due to the
$SO(3)$ symmetry)

\section{Exact Spectrum and Instabilities}
\noindent

The exact spectrum of string theory in the magnetic background
described
in the last section can be computed by solving the associated
conformal
field theory, \cite{magnetic}.
If we call $M^2_{L}$ the eigenvalues of $L_{0}$ and $M^2_{R}$ the
eigenvalues of $\bar L_{0}$ we find
\be
M^2_{L}=-{1\over 2}+{\Q^2\over 2}+\sum_{i=1}^{3}{\Q_{i}^2\over
2}+{(j+1/2)^2-(\Q+I)^2\over
k+2}+E_{0}
+{\left[{(\Q+I)\over \sqrt{k+2}}+eH\right]^2\over 1-2H^2}
\label{k11}\ee
\be
M^2_{R}=-1+{\pq^2\over k_{g}}+{(j+1/2)^2-(\Q+I)^2\over
k+2}+\bar E_{0}+{\left[{(\Q+I)\over \sqrt{k+2}}+eH\right]^2
\over 1-2H^2}
\label{k12}\ee
where, the $-1/2$ is the universal intercept in the N=1 side,
$\Q$ is the spacetime helicity,
$\Q_{i}$
are the internal helicity operators (associated to the internal
left-moving fermions),
$E_{0},\bar E_{0}$ contain the oscillator contributions as well as
the internal
lattice
(or twisted) contributions, and
$j=0,1,2,\cdots,k/2$\footnote{Remember that $k$ is an even integer
for $SO(3)$.}, $j\geq |I|\in Z$.
$\pq$ is the zero mode of the affine current associated to the
relevant
gauge group and $e=\sqrt{2}\pq/\sqrt{k_{g}}$.
There is also the usual GSO projection $\Q+\sum_{i=1}^3 \Q_{i}=$ odd
integer.

We can see here another reason for the need of the SO(3)
projection. We do not want half integral values of $I$ to change the
half-integrality of the spacetime helicity $\Q$.
Since for physical states $M^2_{L}=M^2_{R}$ it is
enough to look
at $M_{L}^2$ which in our conventions is the side that has
$N=1$ superconformal symmetry.

The first observation we can make here is to confirm the existence of
a maximal
magnetic field (\ref{max}) suggested from the effective field theory
analysis.
It is obvious from (\ref{k11},\ref{k12}) that at $H=1/\sqrt{2}$ all
states that couple to the magnetic field become infinitely massive.

It is not difficult to check that spacetime fermions have always
positive mass
square, a property required by unitarity.

For bosons though, states with non-zero helicity can become tachyonic
for some range of values  of the magnetic field.
It can be shown that only helicity-one ($\Q=1$) states can become
tachyonic.
Such states have also $E_{0}=\Q_{i}=0$ and $j=|I|$.
Thus there are instabilities provided
\be
{1\over 1-2H^2}\left({(1+I)\over
\sqrt{k+2}}+eH\right)^2+{(|I|+1/2)^2-(1+I)^2\over
k+2}\leq 0\label{tach}
\ee
and
\be
{1\over 2(k+2)} \leq e^2 \leq 2\label{k20}
\ee
Introducing the mass gap $\mu^2=1/(k+2)$ we obtain tachyonic
instabilities when
\be
H^{\rm crit}_{\rm min}\leq |H| \leq H^{\rm crit}_{\rm max}
\ee
with
\be
H^{\rm crit}_{\rm min}={\mu\over |e|} {1-{\sqrt{3}\over
2}\sqrt{1-{1\over 2}\left({\mu\over e}\right)^2}\over
1+{3\over 2}\left({\mu\over e}\right)^2}
\;\;,\;\;
H^{\rm crit}_{\rm max}={\mu\over |e|}{J+1+ \sqrt{\left(J+{3\over
4}\right)\left(1-2\left(J+{1\over 2}\right)^2{\mu^2\over
e^2}\right)}\over
1+\left(2J+{3\over 2}\right){\mu^2\over e^2}}
\label{q1}
\ee
and
\be
J={\rm integral ~~part~~of~~~}-{1\over 2}+{|e|\over \sqrt{2}\mu}
\ee

We note that for small $\mu$ and $|e|\sim {\cal O}(1)$ $H^{\rm
crit}_{\rm min}$ is of order ${\cal O}(\mu)$.
However $H^{\rm crit}_{\rm max}$ is below $H_{\rm max}=1/\sqrt{2}$ by
an amount of order ${\cal O}(\mu)$.
Thus for small values of $H$ there are no tachyons until a critical
value $H^{\rm crit}_{\rm min}$ where the theory becomes unstable. For
$|H|\geq H^{\rm crit}_{\rm max}$ the theory is stable again till the
boundary $H=1/\sqrt{2}$.
It is interesting to note that if there is a charge in the theory
with the value $|e|=\sqrt{2}\mu$ then $H^{\rm crit}_{\rm
max}=1/\sqrt{2}$ so there
is no region
of stability for large magnetic fields.
For small $\mu$ there are always charges satisfying (\ref{k20}) which
implies that there is always a magnetic instability.
However even for $\mu={\cal O}(1)$ the magnetic instability is
present for standard gauge groups that have been considered in string
model building (provided they  have charged states in the
perturbative
spectrum).

The behavior above should be compared to the field theory behavior
(\ref{class}). There we have an instability provided there is a
particle with $gS\geq 1$. Then the theory is unstable for
\be
|H|\geq {M^2\over |e|(gS-1)}
\ee
where $M$ is the mass of the particle (or the mass gap).
However there is no restauration of stability for large values of
$H$.
This happens in string theory due to the backreaction of gravity.
There is also another difference. In field theory $H_{crit}\sim
\mu^2$
while in string theory $H_{crit}\sim \mu M_{\rm Plank}$ where we
denoted by
$\mu$ the mass gap in both cases.
This is due to the different ways of breaking the gauge symmetry.

A discussion on the flat space limit ($\mu\to 0$) of these solutions
can be found in \cite{magnetic}.

\section{Conclusions and Further Comments}
\setcounter{equation}{0}
\noindent
We have presented a class of magnetic backgrounds in
closed superstrings and their associated instabilities.
Our starting point are superstring ground states with an adjustable
mass
gap $\mu^2$ \cite{kk}.
In such ground states all gauge symmetries are spontaneously broken.

Exact magnetic and gravitational solutions can then be constructed in
such ground states as exactly marginal perturbations of the
appropriate CFTs.
In the magnetic case, there is  a monopole-like magnetic field on
$S^3$.
The gravitational backreaction squashes mildly the $S^3$ keeping
however an $SO(3)$ symmetry.
We have calculated the exact spectrum as a function of the magnetic
field.
The first interesting observation is that, unlike field theory, there
is a
maximum value for the magnetic field $\sim M_{\rm Plank}^2$.
At this value the part of the spectrum that couples to the magnetic
field
becomes infinitely massive.

We find magnetic instabilities in such a background.
In particular, for $H\sim {\cal O}(\mu M_{\rm Plank})$ there is a
magnetic instability,
driven by helicity-one states that become tachyonic.
The critical magnetic field scales differently from the field theory
result, due the different mechanism of gauge symmetry breaking.

We also find that, unlike field theory, the theory becomes stable
again for
strong magnetic fields of the order $\sim {\cal O}(M^{2}_{\rm
Plank})$.

Such instabilities could be relevant in cosmological situations, or
in black hole evaporation.
In the cosmological context, there maybe solutions where one has time
varying
long range magnetic fields. If the time variation is adiabatic, then
there might be
a condensation which would screen and localize the magnetic fields.
Also, instabilities can be used as (on-shell) guides to find the
correct vacuum of string theory.
Our knowledge in that respect is limited since we do not have an
exact description of all possible deformations
of a ground state in string theory.

Another subject of interest, where instabilities could be relevant is
Hawking radiation.
It is known in field theory that Hawking radiation has many common
features with production
of Schwinger pairs in the presence of a long range electric field.
In open string theory it was found, \cite{bp} that this rate becomes
infinite for a
$finite$ electric field , $E_{\rm crit}\sim M^2_{\rm string}$ (unlike
the field theory case) and this behavior is due
to $\alpha'$ corrections. Notice also that in the open string it is
$M_{\rm string}$ and not $M_{\rm Plank}$ that is relevant due to the
absence of gravity.
It would be interesting to see if this behavior persists in the
presence of gravity
(which is absent to leading order in open strings)
by studying the effect in closed strings.
In fact we expect that gravitational effects will be important for
$E\sim M^2_{\rm planck}$. For small $g_{\rm string}$ however,
we can have
$M_{\rm string} << M_{\rm planck}$ so we expect a similar behavior as
in the case of open strings.
It is plausible that similar higher order corrections modify the
Hawking rate
in such a way that macroscopic black hole are unstable in string
theory.
Such a calculation seems difficult to perform with today's technology
but seems crucial to
the understanding of stringy black holes.
\vskip 1cm

\centerline{\bf Acknowledgements}

E. K. would like to thank the organizers for
their warm hospitality. C. Kounnas was  supported in part by EEC
contracts SC1$^*$-0394C and SC1$^*$-CT92-0789.

\end{document}